\documentclass[aps, twocolumn, prl, citeautoscript, superscriptaddress, amsmath, 8pt]{revtex4-2}

\usepackage{graphicx}
\usepackage{gensymb}
\usepackage{color}
\usepackage[dvipsnames]{xcolor}
\usepackage{float}
\usepackage{upgreek}
\usepackage{bm}
\usepackage{amsmath,amssymb}
\usepackage[colorlinks=true,citecolor=blue,linkcolor=blue]{hyperref}
\setcitestyle{super}

\begin{document}

\title{Imaging magnetic switching in orthogonally twisted stacks of a van der Waals antiferromagnet}

\author{Alexander J. Healey} 
\affiliation{School of Science, RMIT University, Melbourne, VIC 3001, Australia}

\author{Cheng Tan} 
\affiliation{School of Science, RMIT University, Melbourne, VIC 3001, Australia}

\author{Boris Gross}
\affiliation{Department of Physics, University of Basel,  Basel, Switzerland }

\author{Sam C. Scholten}
\affiliation{School of Physics, University of Melbourne, Melbourne, VIC 3010, Australia}

\author{Kaijian Xing}
\affiliation{School of Science, RMIT University, Melbourne, VIC 3001, Australia}

\author{Daniel G. Chica}
\affiliation{Department of Chemistry, Columbia University, New York, NY, USA}

\author{Brett C. Johnson}
\affiliation{School of Science, RMIT University, Melbourne, VIC 3001, Australia}

\author{Martino Poggio}
\affiliation{Department of Physics, University of Basel,  Basel, Switzerland }
\affiliation{Swiss Nanoscience Institute, University of Basel, Basel, Switzerland}

\author{Michael E. Ziebel}
\affiliation{Department of Chemistry, Columbia University, New York, NY, USA}

\author{Xavier Roy}
\affiliation{Department of Chemistry, Columbia University, New York, NY, USA}

\author{Jean-Philippe Tetienne}
\email{jean-philippe.tetienne@rmit.edu.au}
\affiliation{School of Science, RMIT University, Melbourne, VIC 3001, Australia}

\author{David A. Broadway}
\email{david.broadway@rmit.edu.au}
\affiliation{School of Science, RMIT University, Melbourne, VIC 3001, Australia}

\begin{abstract}
Stacking van der Waals magnets holds promise for creating new hybrid materials with properties that do not exist in bulk materials.
Here we investigate orthogonally twisted stacks of the van der Waals antiferromagnet CrSBr, aiming to exploit an extreme misalignment of magnetic anisotropy across the twisted interface.
Using nitrogen-vacancy centre microscopy, we construct vector maps of the magnetisation, and track their evolution under an external field, in a range of twisted compensated and uncompensated configurations differing by the number of layers.
We show that twisted stacking consistently modifies the local magnetic switching behaviour of constituent flakes, and that these modifications are spatially non-uniform. 
In the case of compensated component flakes (even number of layers), we demonstrate that the combination of dipolar coupling and stacking-induced strain can reduce the switching field by over an order of magnitude. 
Conversely, in uncompensated component flakes (odd number of layers), we observe indications of a non-zero interlayer exchange interaction between twisted flakes during magnetization reversal, which can persistently modify magnetic order. 
This work highlights the importance of spatial imaging in investigating stacking-induced magnetic effects, particularly in the case of twistronics where spatial variation is expected and can be conflated with structural imperfections. 
\end{abstract}

\maketitle 

Tuning the properties of magnetic materials is a critical challenge for the development of new devices. 
One way of achieving this is to create heterostructures of different magnetic thin films to modulate or combine their properties~\cite{bernevig_progress_2022}. 
However, in this case deleterious interfacial effects are common and it can be difficult to control the direction of anisotropy axes during deposition. 
The bottom-up assembly of van der Waals materials presents an intriguing alternative, opening a toolbox to explore engineering new magnetic interactions and phases without incurring the negatives associated with three dimensional thin films. 
Beyond magnetism, there is a large precedent for utilising twisted stacking and moir\'e patterning~\cite{carr_electronic-structure_2020, ciarrocchi_2022} to alter interlayer electronic interactions~\cite{cao_correlated_2018, cao_unconventional_2018, inbar_quantum_2023, cai_signatures_2023, park_observation_2023, zhang_twist-angle_2020}. 
The intersection of both ideas has seen the exploration of stacking order-induced modification of magnetic properties~\cite{Sivadas_stacking_dependent_2018, Chen_direct_2019, Thiel2019}, and early demonstrations of magnetic moir\'e lattices~\cite{song_direct_2021}. 
The wide range of competing magnetic interactions introduced by altered stacking configurations creates a complex landscape that may allow varied magnetic phases to be sustained and tuned through stimulus or background field variation\,\cite{klein_control_2022}.

\begin{figure*}[ht]
	\centering
    \includegraphics[width=\textwidth]{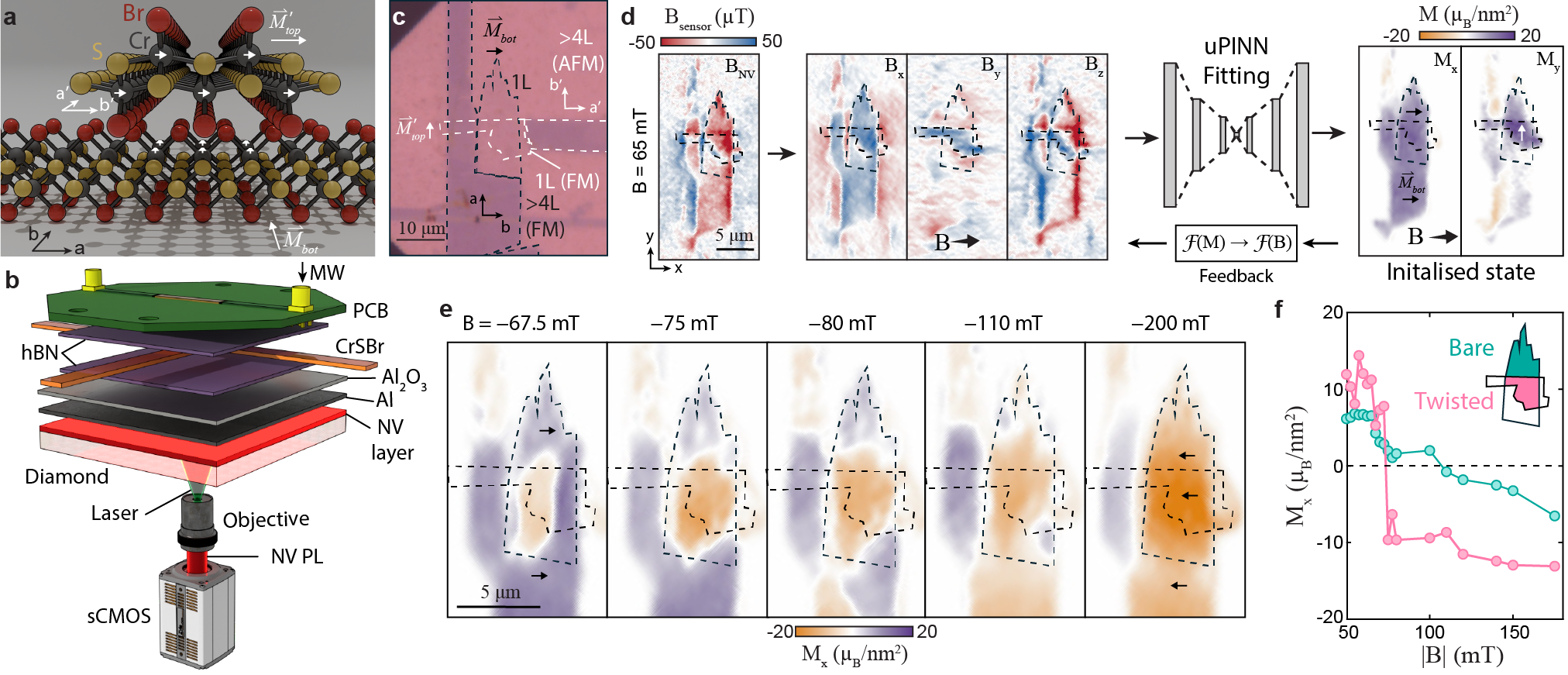}
	\caption{\textbf{FM switching in a twisted FM-FM bilayer.}
        \textbf{a}~Schematic of two monolayers of 90$^\circ$ twisted CrSBr flakes.
        \textbf{b}~Schematic of the widefield NV microscopy setup (see SI for details).
        \textbf{c}~Optical image of an encapsulated 90$^\circ$ twisted stack of two CrSBr monolayers. 
        \textbf{d}~Sketch of measurement and analysis procedure used throughout the paper. A magnetic image recorded through NV magnetometry (far left) is converted to components $B_x$, $B_y$, $B_z$ through Fourier methods. 
        These component images are then fed into an untrained physically informed neural network (uPINN) to reconstruct the easy-plane magnetisation, $M_x$ (left) and $M_y$ (right). 
        \textbf{e}~A series of $M_x$ images (monolayers highlighted with dashed lines) collected after initialising along the x-direction as shown in \textbf{d} and then increasing $B$ in the opposite direction, showing the FM flipping transition. Domains are seen to nucleate in the twisted region before spreading to the rest of the flake.
        \textbf{f}~$M_x$ hysteresis loop for the twisted and bare regions indicated taken from \textbf{e}~and extended data (see SI). 
     }
\label{fig intro}
\end{figure*}

Spatially varying interactions (e.g. exchange) can be present in twisted stacks related to their Moir\'e superlattices whose effects are often observed at small twist angles (or large superlattices).
Alternatively, interlayer interactions can also be controlled by much larger twist angles, up to an extreme of 90$\degree$.
In this case, we may again expect modulated interlayer coupling due to the new stacking configuration\,\cite{kapfer_programming_2023}, but, where strong uniaxial anisotropy exists within the plane, a significant variation in easy magnetisation axes between the two layers is also introduced. 
Importantly, in the extreme (orthogonal) configuration, any Heisenberg-type interlayer exchange that exists between twisted layers will be ``turned off" when spins are pinned along their respective easy axes, only becoming active when spins begin to rotate, such as in a domain wall or across a gradual spin canting transition.
However, orthogonally twisted interfaces [sketched in Fig.~\ref{fig intro}\textbf{a}] may introduce non-Heisenberg exchange pathways [e.g. Dzyaloshinskii–Moriya interactions (DMI)], where any nonzero interactions may enable new magnetic phases or behaviours to be sustained at the twisted interface, perhaps occurring preferentially across switching transitions where spins realign or domain walls form.

To explore this model of orthogonal anisotropy, this work investigates orthogonally twisted stacks of the magnetic semiconductor CrSBr~\cite{telford_layered_2020, ziebel_crsbr_2024}, an A-type antiferromagnetic (AFM) semiconductor~\cite{telford_designing_2023, klein_control_2022}, where each layer is ferromagnetic (FM), with strong uniaxial in-plane anisotropy giving an easy axis along the crystal $b$-direction. 
Recent work using ``orthogonally twisted bilayers" demonstrated complex magnetic switching behaviour via magneto-resistance measurements~\cite{boix-constant_multistep_2023}, and this twisted magnetic phased space was shown to be useful as a mechanism for magnetic memory~\cite{Chen2024}.
While these electrical measurements (averaged over the whole device) are suggestive of complex domain wall movement related to the twisted stacking, their microscopic origin remains ambiguous. 
In particular, an open question is whether these steps are tied to intrinsic exchange interactions in the orthogonal structure or extrinsic strain and stacking defects. 
Understanding the relative contributions of these effects is critical in the future design of related devices.
As such, it is critical to understand how these magnetic structures form on the nano/micron scale, especially given that stacking disorder is a known problem for vdW assemblies~\cite{li_towards_2024}.

Here, we image the magnetic switching behaviours in orthogonally twisted CrSBr stacks using a cryogenic widefield nitrogen-vacancy (NV) centre microscope\,\cite{Broadway2020} [Fig.~\ref{fig intro}\textbf{b}], which offers rapid, quantitative magnetic imaging at a sub-micron resolution~\cite{Scholten_widefield_perspective_2021}. 
Visualising domain formation and propagation allows us to confirm the behaviours observed originate in the twisted region and assess the impact of conflating factors such as stacking-induced strain. 
Furthermore, the technique allows the extraction of vector information~\cite{Casola_cmp_review_2018}, yielding insight into the interactions behind the global properties of devices by separating the contributions from the constituent layers. 
Using these techniques we explore a range of twisted samples made of different numbers of layers: 1L(FM)-1L(FM), 2L(AFM)-2L(AFM), 3L(FM)-2L(AFM), where the magnetic ground state in zero magnetic field is indicated in parentheses (FM means there is one uncompensated magnetic layer, whereas AFM is fully compensated).
The interlayer dipolar coupling in these structures varies from strong (FM-FM) to vanishing (AFM-AFM), highlighting the role of interlayer exchange interaction in modifying the magnetic behaviour of twisted stacks. 

We begin with the orthogonally twisted bilayer (1L-1L) configuration investigated by Boix-Constant \textit{et al.}~\cite{boix-constant_multistep_2023}, see optical image Fig.~\ref{fig intro}\textbf{c}. 
In general, both constituent monolayers are expected to exhibit net magnetisation lying along their respective easy axes (which we refer to as $b$ and $b'$, see Fig.~\ref{fig intro}\textbf{a}), resulting in complex magnetic stray field images.
To facilitate interpretation, we employ an untrained physically informed neural network approach~\cite{Dubois_reconstruction_2022} [Fig.~\ref{fig intro}\textbf{d}]; first we convert the single-sensor magnetic field image into the field components (e.g. $B_x$, $B_y$, $B_z$)~\cite{Casola_cmp_review_2018} and then the neural network reconstructs the source magnetisation within the easy plane by fitting the reconstructed magnetic fields~\cite{tschudin_imaging_2024}. 
Through this analysis, we can obtain spatially resolved information about magnetisation direction and magnitude, details which are typically invisible to transport-based measurements. 

To assess the evolution of the magnetisation we first initialise one of the flakes with a large positive field ($B = 260$\,mT), shown in Fig.~\ref{fig intro}\textbf{d}.
Next, we apply negative field pulses of increasing magnitude to flip the magnetization and measure at a low, non-invasive measurement field of $B \approx 5$~mT.
We note that the magnetic field is applied approximately along the desired $b$-axis of the targeted flake (matching the $x$-axis in Fig.~\ref{fig intro}\textbf{d}) with a $z$-component (polar angle $\theta = 54.7^\circ$) which is required for matching the orientation of the NV sensor.
Using this protocol, we increased the $B$ field along the $b$-axis of the ``aligned'' flake, corresponding to the $a^\prime$-axis of the ``non-aligned'' flake [Fig.~\ref{fig intro}\textbf{e}]. 
We observe that domain switching within the aligned flake preferentially occurs in the twisted region before propagating step-wise through the rest of the flake. 
In Fig.~\ref{fig intro}\textbf{f} we summarise the results of the sweep by recording the average reconstructed magnetisation values in the ``twisted" and ``bare" monolayer regions, where the twisted region experiences a sharp transition at $B=-75$\,mT and the bare region experiences a slower smooth transition.

An intriguing interpretation of the reduced switching field within the twisted region is that an interlayer exchange interaction lowers the energy for domain wall nucleation. 
Indeed, this interpretation is supported by the observation that the domain border in Fig.~\ref{fig intro}\textbf{e} is collinear with the b' axis. 
Still, we must consider extrinsic effects.
We do not observe weakened magnetism (a lower magnetic moment or increased inhomogeneity) within the twisted region that could indicate structural damage caused by the stacking.
Local dipolar fields from defects below our spatial resolution could result in a local reduction in the switching energy cost but would not preferentially affect the twisted region over others.
Likewise, stacking-induced strain would unlikely be purely limited to the twisted region as each layer/process will introduce inhomogeneous strain including the double encapsulation of hBN and thus cannot explain the overall difference in magnetic behaviour.
The role of edge fields cannot be discounted given the shape and size of flakes in the twisted stack, and domain nucleation does correlate with the edge of the orthogonal flake.
However, an exchange interaction could produce a local reduction in the energy cost for nucleating a domain wall with a helicity that matches the sign of the exchange interaction with the orthogonal flake.
These results alone support the suggestion from previous work\,\cite{boix-constant_multistep_2023} that a nontrivial interaction exists between the two flakes in such a configuration, and we will further explore this hint of exchange-driven FM reversal later.

\begin{figure}[t]
	\centering
    \includegraphics{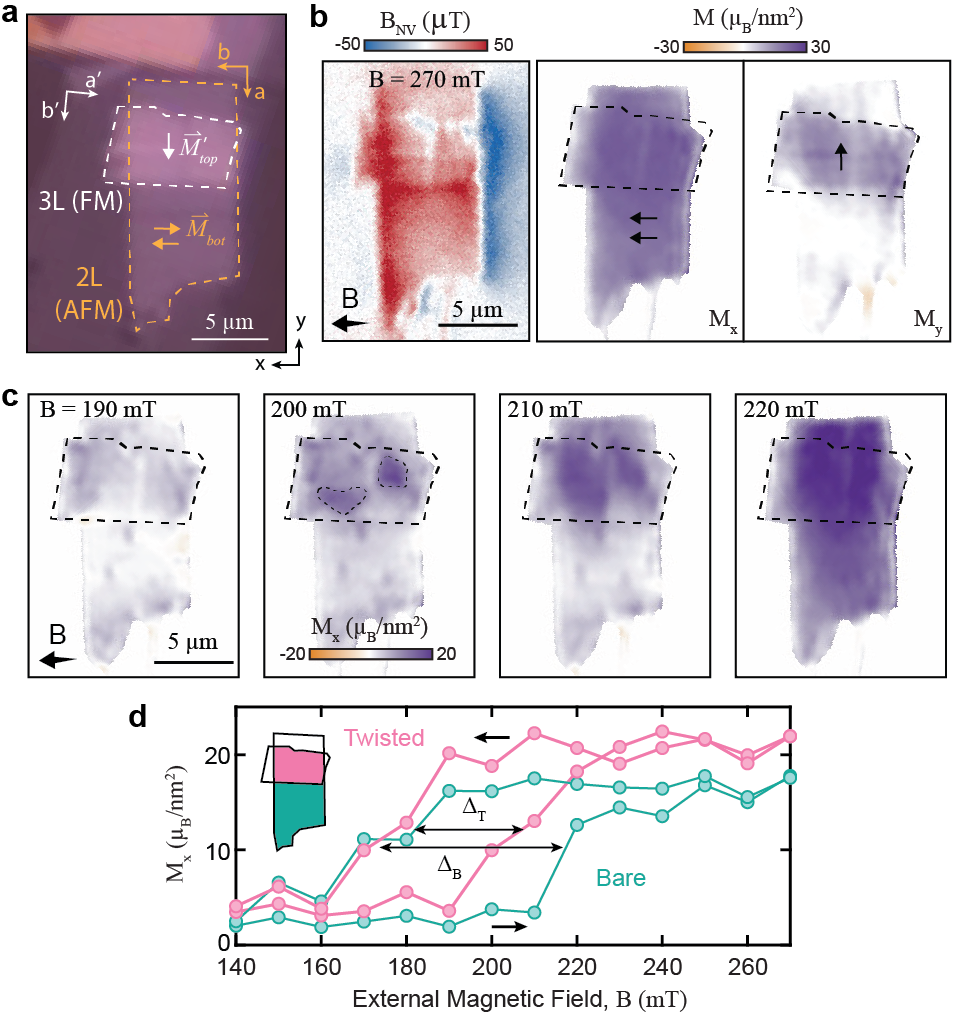}
	\caption{\textbf{AFM switching in an AFM-FM twisted stack.}
        \textbf{a}~Optical image of the sample with the two flakes outlined.
        \textbf{b}~Example magnetic image recorded at an applied field 270~mT along $b$, past the AFM-to-FM phase flip transition of the aligned flake, and reconstructed magnetisation components $M_x$ and $M_y$. 
        \textbf{c}~Series of images collected upon increasing fields. Phase flips in the AFM flake are seen to nucleate in the twisted region. 
        \textbf{d}~$M_x$ hysteresis taken from the series in \textbf{c}~and extended data (see SI) from the regions indicated. The different hysteresis gaps, $\Delta_B = 51(28)$\, mT for the bare region, $\Delta_T = 32(15)$\, mT for the twisted region, were determined by a Lorentzian fit of the difference between sweep directions (see SI).
     }
\label{fig2}
\end{figure}


\begin{figure}[t]
    \centering
    \includegraphics{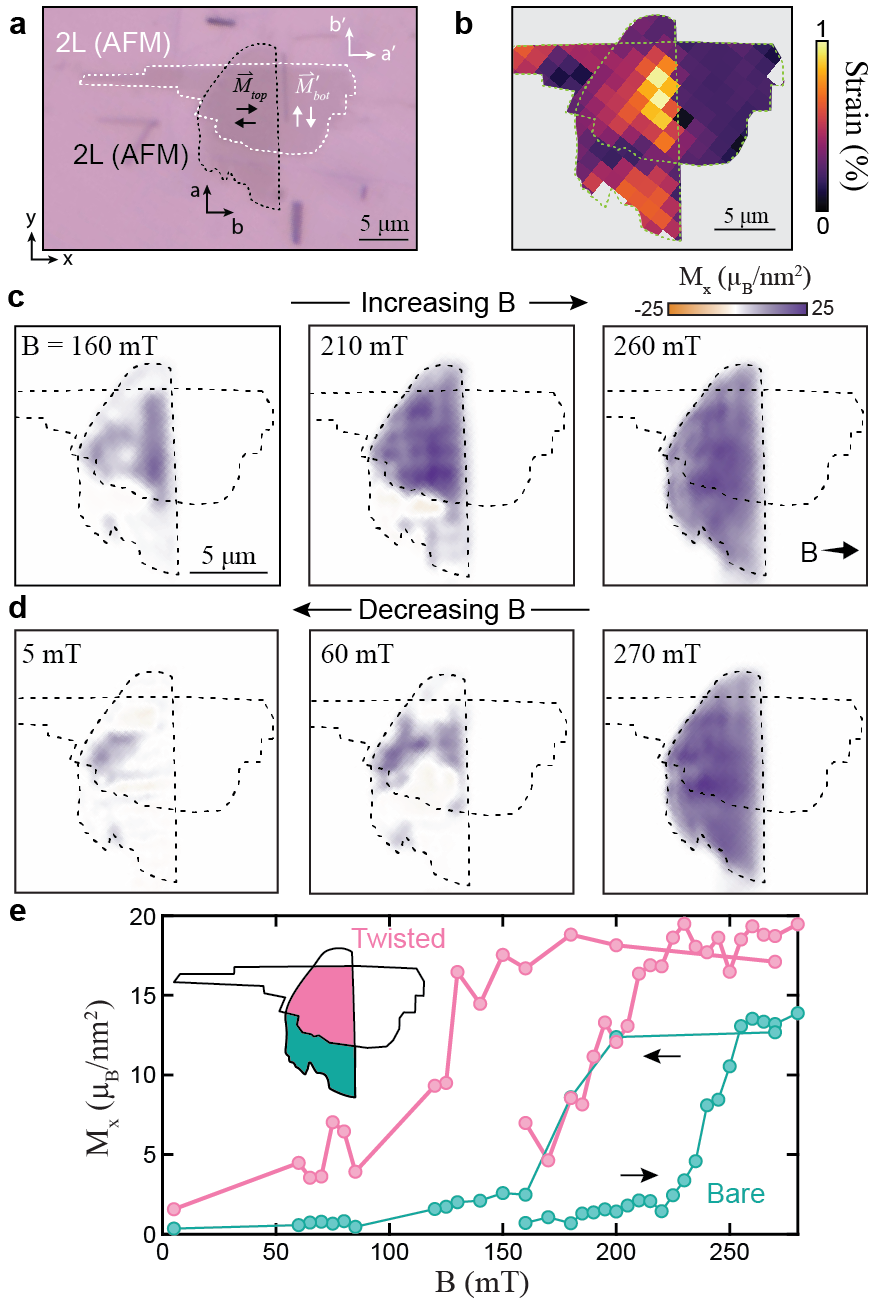}
    \caption{ \textbf{AFM switching in an AFM-AFM twisted stack}. 
        \textbf{a}~Optical image of a twisted 2L(AFM)-2L(AFM) stack.
        \textbf{b}~Strain in the twisted stack measured through shifts in the $P3$ Raman line (see SI). 
        \textbf{c}~In-plane magnetisation images with an increasing field across the AFM-to- FM transition.
        \textbf{d}~In-plane magnetisation images with a decreasing field across the FM-to-AFM transition, with a persistent FM domain at $B=5$~mT.
        \textbf{e}~Magnetic hysteresis loop for both the twisted and bare regions as indicated.
    }
\label{Fig afm}
\end{figure}

We now investigate whether the interaction between twisted layers can be similarly exploited to locally induce new switching behaviours in AFM flakes comprising an even number of layers, which are appealing to incorporate into devices since they naturally minimise stray field~\cite{Chen2024}. 
We start with a 3L(FM)-2L(AFM) stack shown in Fig.~\ref{fig2}\textbf{a}, and focus on the metamagnetic phase transition between the AFM 2L configuration (i.e. the zero-field ground state) and a forced FM configuration observed when appreciable fields are applied along the $b$-axis (overcoming the interlayer AFM coupling to switch the anti-aligned layer). 
Using the directional information present in our measurements, we separate the magnetisation belonging to each flake, shown in the forced FM case in Fig.~\ref{fig2}\textbf{b}. As before, the magnetisation is relatively uniform in the saturated state, indicating good overall structural integrity of the flakes despite the stacking process.  
To directly probe the magnetism across the transition of interest we adapt our experimental protocol to measure at a given field $B$ rather than cycling between separate measurement and pulse fields (we retain this measurement scheme for the remainder of the manuscript). 

Starting from the compensated configuration of the 2L flake (zero net $M_x$) and sweeping through the AFM-to-FM phase transition [Fig.~\ref{fig2}\textbf{c}], we again see that the phase switching selectively nucleates in the twisted region before spreading to the entire flake. 
The average $M_x$ magnetisation within the twisted and bare regions [Fig.~\ref{fig2}\textbf{d}] shows that the hysteresis loop is narrower ($\Delta_T = 32(15)$\,mT vs $\Delta_B = 51(28)$\,mT) for the twisted region.
This is accompanied by a stronger saturation magnetisation in the twisted region, which can be explained by a partial canting of the 3L flake towards the applied field. 
In the Supplementary Information we show that the 3L flake is similarly hysteretic across the bilayer spin-flip transition. In principle, this observation supports the presence of an interlayer exchange interaction across the twisted interface. Here, this interaction gives rise to a collective magnetic transition extending across both flakes, in which the spin-flip of the AFM bilayer is coupled to a reorientation of the FM 3L flake. These results are in qualitative agreement with a micromagnetic simulation incorporating an interflake exchange coupling.

Beyond this proposed interlayer exchange interaction, AFM switching might also be affected by contributions from dipolar coupling and strain. To disentangle these effects, we next image an orthogonally twisted 2L(AFM)-2L(AFM) stack [Fig.~\ref{Fig afm}\textbf{a}] that minimises the net dipolar fields emitted in the magnetic ground state.
To quantify strain, we perform Raman spectroscopy [Fig.\,\ref{Fig afm}\textbf{b}] and use the known relationship between strain and the $P3$ Raman line~\cite{Cenker2022} which is separated from the layer-dependent shifts of other Raman lines~\cite{torres_probing_2023}.
Our twisted stack shows significant strain in the twisted region, presumably from a combination of a frustrated stacking pattern and the stacking process itself.
Increasing $B$ across the AFM-to-FM transition of the top flake, we observe that FM domains nucleate in the twisted region first (Fig.\,\ref{Fig afm}\textbf{c}). Given the correlation with the higher strain region, we cannot rule out that the reduced switching field in this sample could be dominated by local strain. Note that domain nucleation in the bottom flake ($a^{\prime}b^{\prime}$) does not appear to correlate with strain (see SI), which could indicate that the strain is mostly contained in the top flake ($ab$).  
Likewise, upon decreasing $B$ the flake switches back to the AFM configuration at significantly lower fields in the twisted region compared to bare 2L region (Fig.\,\ref{Fig afm}\textbf{d}), by an average of 50-100 mT as shown in the hysteresis loop (Fig.\,\ref{Fig afm}\textbf{e}).
Interestingly, in Fig.\,\ref{Fig afm}\textbf{d} we observe strong FM domains persisting down to an applied field of just $B = 5$~mT (over an order of magnitude below the bare case), which may be attributed to localised structural defects or magnetic interactions that weaken the AFM interlayer coupling within the 2L flake. 

To develop a deeper understanding of the origin of the spatial variation of the switching field, we performed micromagnetic simulations\,\cite{Vansteenkiste2014, Leliaert2017, Exl2014} of this twisted stack using the known properties of CrSBr\,\cite{ziebel_crsbr_2024} in the same fashion as in Ref.\,\cite{tschudin_imaging_2024} (see SI for more details).
The simulations suggest that the reduced switching field and extended hysteretic gap are governed by the combination of strain-induced softening of the interlayer exchange interaction\,\cite{Cenker2022} and dipolar coupling between the twisted flakes. 
As such, we conclude that unlike the FM switching case of Fig.~1 which suggests an interlayer exchange between the twisted layers, strain and dipolar coupling are likely to play a dominant role in explaining the AFM switching results.


\begin{figure}[htb]
	\centering
    \includegraphics[scale=1]{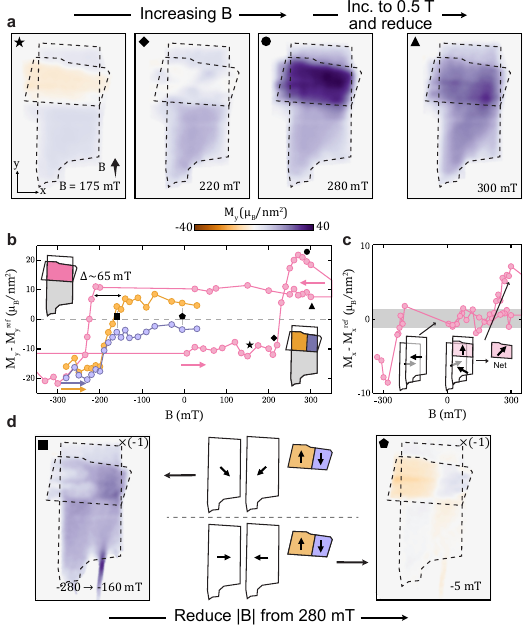}
	\caption{\textbf{FM switching in an AFM-FM twisted stack}. 
         \textbf{a}~Series of reconstructed $M_y$ images collected with increasing applied field along the $y$-axis for the AFM-FM stack shown in Fig.~\ref{fig2}\textbf{a}.
         \textbf{b}~Summary of the full measurement series from which the images in \textbf{a} are sourced, showing reconstructed $M_y$ magnetisation in the twisted region (pink region in left hand schematic). The average magnetisation is normalised against the average in the bare bilayer ($M_y^{\rm ref}$ from the grey region in schematics) so as to subtract the contribution of bilayer canting. Purple and orange traces show a separate set of data where the field is reduced from a maximum value of 280~mT, corresponding to the peak of the ``overshoot" and partial 3L flipping. These traces show the average magnetisation in the respective regions in the right hand schematic. Symbols correspond to the images from \textbf{a} and \textbf{d} as indicated.
         \textbf{c}~Average reconstructed $M_x$ magnetisation from the same data set. Inset schematics show the inferred evolution of the bilayer sublattices. 
         \textbf{d}~Exemplary images from the sweep obtained upon reducing from a maximum of 280~mT with schematics of the inferred magnetic orientation of the constituent sublattices (middle). Magnetisation values have been inverted here to facilitate easier comparison with \textbf{a}.  
         }
\label{Fig domains}
\end{figure}

Finally, we return to FM switching, by considering the 2L(AFM)-3L(FM) stack from Fig.~\ref{fig2} to investigate the effect of an orthogonally twisted AFM flake on the flipping transition of the FM 3L. 
In this case, the range of interactions introduced by the orthogonal stack may be expected to complicate domain nucleation and reorientation (as we saw in Fig.~\ref{fig intro}), allowing states not supported in the pristine material to be maintained even down to zero field.
Magnetic images recorded across the 3L flipping transition are shown in Fig.~\ref{Fig domains}\textbf{a}. The reconstructed $M_y$ magnetisation in the bare 2L region is seen to gradually cant towards the applied field, while domains form around 220~mT in the 3L as its magnetisation switches. The transition appears complete at 280~mT. However, curiously, we see that upon increasing the applied field further (500 mT) and reducing back to a field strength accessible to our NV microscope (300~mT) the magnitude of the magnetisation is reduced.  

The full set of data from which these images were taken is summarised in Fig.~\ref{Fig domains}\textbf{b}. Due to the canting of the 2L flake, we normalise the average $M_y$ magnetisation measured in the twisted region by subtracting the average in the bare region (the pink and grey regions in the inset schematic, respectively). In the case of the twisted flakes being totally uncoupled, we would expect to recover a square hysteresis loop typical of a hard ferromagnet. Instead, we observe an approximately twofold increase in net magnetisation as the 3L flips, before it reduces back to the standard monolayer magnetisation as the field is increased further. This region of excess $M_y$ magnetisation, which we dub an ``overshoot" due to its tendency to return to the monolayer value at higher fields, is accompanied by a spike in net $M_x$ magnetisation, which is otherwise zero within error (Fig.~\ref{Fig domains}\textbf{c}). This directionally resolved data implies that the excess magnetisation originates from the reorientation of spins within the bilayer at the twisted interface towards its $a'$ axis, leading to an increase in $M_y$ and a loss of full $M_x$ compensation. Simplified sketches of the proposed magnetic structure in the 2L flake before and after this transition are inset in Fig.~\ref{Fig domains}\textbf{c}.

We can see concretely that the dominant component of the overshoot magnetisation originates in the 2L flake by reducing the applied field from a maximum of 280~mT, just before the peak in net magnetisation, to near zero field. This data is plotted as the purple and orange points in Fig.~\ref{Fig domains}\textbf{b}, corresponding to the average normalised magnetisation in the regions highlighted in the rightmost inset schematic.  
As the field is lowered, two distinct, oppositely oriented domains emerge (see exemplary images in Fig.~\ref{Fig domains}\textbf{d}), indicating that the 3L flip is not complete until we move past the overshoot.
We attribute the excess magnetisation (with a hysteretic width $\Delta \sim 65$~mT) to a partial 2L flipping transition similar to that observed in Fig.~\ref{Fig afm}. 
Altogether, the data in Fig.~\ref{Fig domains} provides further evidence of FM exchange interactions across the twisted interface. 
The co-occurrence of the bilayer reorientation with the 3L flipping transition is suggestive of the overshoot magnetisation being driven by the formation of domain walls within the 3L flake.
We propose that the spin alignment within domain walls formed in the 3L flipping transition, which would align to the $b$-axis of the orthogonal flake, may locally enhance the coupling between the layers and result in additional magnetic alignment of the orthogonal flake. For instance, in line with the simulations from Boix-Constant~\textit{et al.}\cite{boix-constant_multistep_2023}, a significant DMI could drive spins in the bilayer towards the intermediate $a$-axis. Once the 3L magnetisation saturates, the magnetic frustration is broken and the bilayer's evolution with field proceeds as a smooth canting as if it were isolated from the 3L flake.

Utilising widefield NV centre microscopy we have explored the evolution of magnetisation in orthogonally twisted CrSBr stacks. 
The presence of two distinct easy axes in the constituent flakes in this hybrid configuration, combined with the non-trivial interaction between layers at the twisted interface allows diverse magnetic textures to be sustained. 
Crucially, the vector information inherent in our measurements can isolate the behaviours of individual flakes based on their respective easy axes in the regime where they are not strongly coupled, and evidence a realignment of spins at the twisted interface where there is strong coupling.
We have found that metamagnetic phase transitions from AFM to FM can be locally dominated by strain, which is important for future transport devices that will need to either have extremely uniform flakes or be on a small enough scale that strain is near uniform across the device junction. 
In contrast, we found that FM domain reversal enhances interlayer exchange coupling during this transition, inducing magnetic reorientation in the non-aligned flake. 
Given these findings, spatial imaging of magnetisation in twisted devices with smaller twist angles and eventually moir\'e lattices, where interlayer exchange interactions are expected to be more significant, is important to develop a robust understanding of magnetic textures in this material. 

\textbf{Acknowledgments}
This work was supported by the Australian Research Council (ARC) through grants FT200100073, DE230100192, and DP220102518. 
Synthesis of the CrSBr crystals was funded by the Columbia MRSEC on Precision-Assembled Quantum Materials (PAQM) under award number DMR-2011738 and the Air Force Office of Scientific Research under grant FA9550-22-1-0389.

\textbf{Author Contributions}
AJH, SCS, and DAB took the NV measurements under the supervision of J-PT and DAB.
MEZ, DGC, and XR supplied the bulk crystals, while CT and KX fabricated the samples, with Raman imaging by BCJ and KX.
AJH, BCJ, and JPT produced the diamond sample. 
DAB performed the Neural network reconstruction. 
BG and MP supplied micromagnetic simulations. 
AJH and DAB wrote the manuscript with input from all authors. 

\textbf{Data Availability} The data supporting the findings of this
study are available from the authors upon reasonable request.

\textbf{Competing Interests Statement} 
The authors declare no competing interests

\bibliographystyle{naturemag}
\bibliography{bib.bib}

\end{document}


\title{Supplementary Information: Imaging local anisotropy modifications from twisted exchange interactions}

\maketitle

\tableofcontents

\section{Experimental details}
\subsection{Experimental Setup}
The experimental setup is illustrated in the main text Fig.\,1(b).
The cryogenic widefield NV microscope was integrated into a closed-cycle cryostat with a base temperature of 4K (AttoDry1000) with a 1-T superconducting vector magnet (Cryomagnetics).
Optical control was performed with a 532~nm continuous wave laser (Laser Quantum Ventus 1~W, coupled to a single-mode fibre and a 60~mm collimation lens at the output of the fibre was used to adjust the beam size at the sample.
The laser was controlled by an acousto-optical modulator (AAOpto MQ180-G9-Fio) and is directed into the cryostat with a dichroic beam splitter, where it passes through a low-temperature microscope objective (Attocube LT-APO/VISIR/0.82).
The NV PL is collected by path which contains a 4f lens configuration and is separated by the dichroic mirror and a 731/137~nm band pass filter, before being focused (300~mm tube lens) onto a water cooled sCMOS camera (Andor Sona). 

The microwave control was performed with a signal generator (Rohde \& Schwarz SMB100A), a switch (Mini-Circuits ZASWA-2-50DR+), and a 50~W amplifier (Mini-Circuits HPA-50W-63), which is directed into the cryostat and through a custom made printed circuit board containing a straight waveguide with a width of 1~mm. 
The measurements were controlled and synchronised with a pulse pattern generator (SpinCore PulseBlasterESR-PRO 500 MHz).

\subsection{Diamond Samples}
The diamond samples were 4.4$\times$4.4~mm diamond slabs with a thickness of 50~$\mu$m (DDK), with both sides polished to a roughness of $<5$\,nm.
The diamonds were implanted with C atoms at 100\,keV, with fluence 1e12~cm$^{-2}$ and were cut into 2.2$\times$2.2~mm chips.
To form the NV layer the diamonds were annealed in a vacuum of $\approx 10^{-5}$ Torr following sequence~\cite{Tetienne2018a}: at 6 h at 400$^\circ$C, 6 h ramp to 800$^\circ$C, 6 h at 800$^\circ$C, 6 h ramp to 1100$^\circ$C, 2 h at 1100$^\circ$C, 2 h ramp to room temperature. After annealing the plates were acid cleaned (15 minutes in a boiling mixture of sulphuric acid and sodium nitrate). Under the experimental conditions used these diamond sensors have sub-10~$\mu$T/Hz$^{1/2}$ sensitivity per diffraction-limited pixel~\cite{Healey2020}.

An Al/Al$_2$O$_3$ (80/80 nm) layer was deposited onto the surface of the diamond sensor chips to protect the CrSBr samples from laser illumination and to maximise the laser intensity at the NV layer.

\subsection{Sample fabrication}
\noindent\textit{Synthesis of CrSBr bulk crystals:}
Large single crystals of CrSBr were grown using a chemical vapour transport reaction described in Scheie et al.~\cite{scheie_spin_2022}.

\noindent\textit{Fabrication of twisted CrSBr samples:}
CrSBr nanoflakes, consisting of 1-3 layers, were obtained by mechanically exfoliation from high-quality single crystals. These nanoflakes were placed onto 285~nm SiO$_2$/Si substrates and examined under optical microscopy to determine the number of layers. Typically, CrSBr nanoflakes have a ribbon shape, with the long direction corresponding to the a-axis and the short direction to the b-axis. Nanoflakes in proximity to the selected one were also used as references to verify the crystalline orientation. After determining the flakes and crystalline orientations the heterostructures were assembled using a standard pick-up method with PDMS/polycarbonate (PC). Subsequently, the assembled heterostructure, with PC on top, was transferred onto a diamond substrate. A fast-dried PMMA layer was then spin-coated onto the top of the PC to encapsulate the entire device. This final step also prevents contamination of the heterostructure by avoiding the use of solvents to remove the PC layer. All procedures were performed in an argon glove box with H$_2$O and O$_2$ levels below 0.1~ppm.

\subsection{Measurement sequence}
\begin{figure}[htb]
    \centering
    \includegraphics{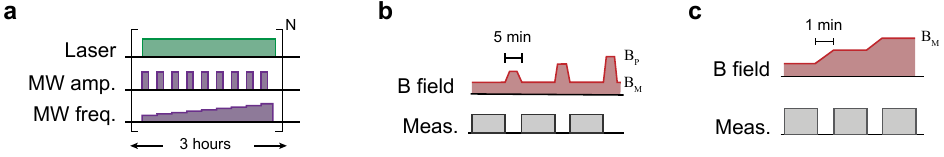}
    \caption{Control sequences used.
    \textbf{a} NV-control and readout sequence where the NV spins are illuminated with a CW laser and the RF field has an amplitude modulation for normalisation and is swept over a range of frequencies. 
    \textbf{b} Magnetic control sequence used for the 1L$\times$1L sample shown in Fig.1 of the main text. The flake is initialised in a new magnetic state by pulsing the magnetic field up to a value $B_p$ and then reduced to a near zero magnetic field for NV measurements.
    \textbf{c} Magnetic control sequence used in all other samples, where the measurement and setting fields were identical. }
    \label{fig:seq}
\end{figure}

Magnetic images were taken using continuous wave optically detected magnetic resonance (CW-ODMR). As depicted in Fig.~\ref{fig:seq}(a), a single measurement sweep over $n$ chosen MW frequencies takes place over $2n$ camera exposures, with every second being a ``MW off" reference measurement. The laser is on throughout and the on/off MW cycle acts as an amplitude modulation to filter out drifts in (for example) laser power. The sequence is then repeated $N<1000$ times (taking three hours or less) to acquire the desired SNR. 

Although optimum sensitivity is in principle obtained by probing a single frequency, since widefield measurements probe a large field of view over which the NV ODMR line shape can vary with changing MW and laser power, capturing the full (or majority) of the resonance linewidth is incentivised to make the measurement robust against these inhomogeneities. Crystal strain within the diamond sensor, by changing the NV zero field splitting frequency, can also inhibit accurate magnetic mapping. Strain effects can be normalised out by taking a reference two-peak measurement, recording maps of the frequencies $f_+$ and $f_-$ and taking the sum ($D = \left( f_+ + f_-\right)/2$). In our measurements the strain variation was negligible compared to the strength of the magnetic signals investigated. 

Two schemes for acquiring sets of data were utilised in this work. Firstly, for Fig~1 of the main text, we used the sequence depicted in Fig.~\ref{fig:seq}(b). In this case, a single low field measurement field $B_M \approx 5$~mT is held constant while a probe field $B_p$ is varied between measurements. This scheme has the advantage of accessing any $B_p$ within the range of the vector electromagnet (up to 1~T in any direction), including those that cannot be measured well using NV microscopy. This approach measures the remanent magnetisation and was appropriate for the FM-FM stack in allowing us to assess whether domains that form at higher fields are stable towards zero field.

For Figs.~2-4 of the main text we preferred the approach sketched in Fig.~\ref{fig:seq}(c), where $B_M$ also acts as the probe field. This sequence is appropriate for stacks including AFM flakes which have fully compensated magnetisation at low fields and directly probing the magnetic behaviours across given transitions is important. 

\section{Analysis}

\subsection{Reconstruction of Magnetisation}
The reconstruction of the magnetisation from the magnetic field was performed by using a physically informed neural network following the method of Dubois et al.\,\cite{Dubois_reconstruction_2022} and modified to fit both the $M_x$ and $M_y$ simultaneous\,\cite{tschudin_imaging_2024}.
Briefly, our magnetic field images are transformed into $B_{xyz}$ using the well known Fourier transformation method~\cite{Casola_cmp_review_2018}.
These three magnetic field components are then fed into a U-net convolutional neural network that splits into two channels at the central node to allow for a sufficient deviation of the reconstructions. 
The output magnetisation images have a mask applied to zero magnetisation contribution outside the flake and are transformed back into magnetic fields using the well-posed transformation.
The mean-square error of the transformed magnetic field and the experimental magnetic fields are computed and the neural network weights are updated accordingly. 

We note that we have a systematic reduction in the magnetisation due to a non-trivial point spread function that results in an underestimation of the magnetic field\,\cite{scholten_aberration_2022}.
However, this reduction is consistent across the images and thus does not change the interpretation of the magnetisation dynamics.

\subsection{Hysteric gap determination}
To obtain a more reliable estimation of the central magnetic field and the hysteresis gap, we take the difference in the sweep directions and fit this transition with a Lorentzian function. The twisted region has a smaller width $\Delta_T = 32(17)$\,mT ($\Delta_B = 51(25)$\,mT) and is centred at a slightly lower magnetic field $B^\prime_T = 191(4)$\,mT ($B^\prime_B = 195(5)$\,mT). 

\begin{figure}[htb]
    \centering
    \includegraphics[width=0.3\linewidth]{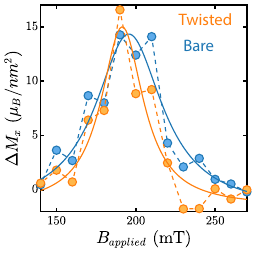}
    \caption{Difference in the magnetic hysteresis for the twisted (orange) and bare (blue) sections of the bilayer flipping with a twisted FM interaction shown in Figure 2 of the main text and Lorentzian fits (solid lines).}
    \label{fig:hyst diff}
\end{figure}

\section{Micromagnetic simulations}
To gain qualitative insight into the importance of various parameters on the behaviours observed, we carried out micromagnetic simulations of some of the scenarios investigated in the main text. The simulations are based directly on the model developed in Ref.~\onlinecite{tschudin_imaging_2024}, and we use the same set of parameters unless stated differently. In brief, black and white images of each layer that are deduced from optical images of
the flake are used to define the geometry. 
The layered structure of CrSBr is then mimicked at hand of the finite difference mesh by setting the cell thickness to the thickness of one
CrSBr layer. Exchange coupling between layers is set to a negative, small fraction of the inplane exchange coupling, facilitating a-type anti-ferromagnetic ordering. Small AFM or FM defects within a stack are used as nucleation points for the respective domains.

To simulate the 1L(FM)-2L(AFM) stack the finite element mesh is set to a size of 1200 x 600 x 3 cells with corresponding cell sizes of \SI{4}{\nano\meter} x \SI{4}{\nano\meter} x \SI{0.8}{\nano\meter}, respectively. This is significantly scaled down to obtain reasonable computation times. The exchange coupling between the two flakes is set to the same value ($-0.008 \cdot A_{ex}$, with $A_{ex}$: inplane exchange stiffness) as the interlayer coupling within a flake. No strain related spatial variation of the intralayer exchange coupling is used in these simulations.
Results of this simulation are shown in Fig.~\ref{fig:AFMFM_sims}, and discussed in the corresponding section.

\begin{figure}[htb]
    \centering
    \includegraphics[width=0.6\linewidth]{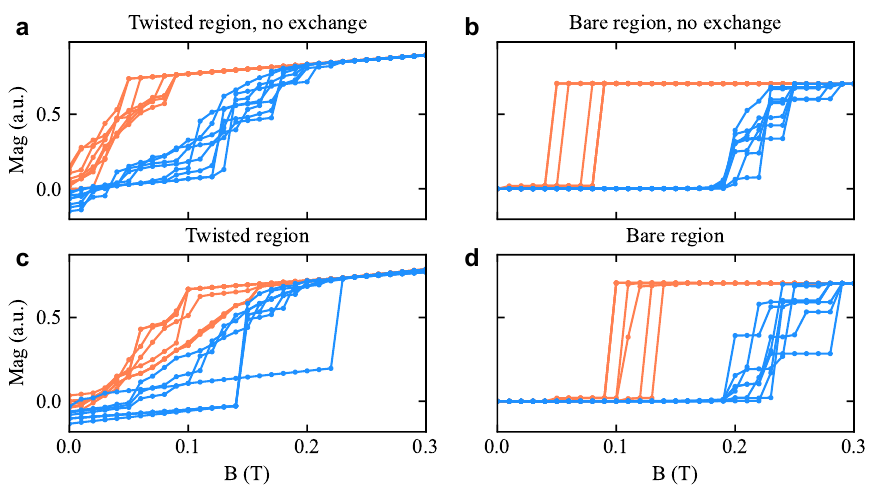}
    \caption{Micromagnetic simulations of the 2L-2L stack (Fig. 3 in the main text). 
    The average magnetisation within the twisted (left; \textbf{a}, \textbf{c}) and bare twisted (right; \textbf{b}, \textbf{d}) regions for multiple simulations both without (top; \textbf{a}, \textbf{b}) and with (bottom; \textbf{c}, \textbf{d}) interlayer exchange coupling between the flakes. Sweeping increasing the applied field in the positive direction are coloured in blue while decreasing sweeps are coloured in orange.}
    \label{fig:magsim}
\end{figure}

The 2L(AFM)-2L(AFM) stack is simulated with a mesh size of 1200 x 1000 x 4 cells and corresponding cell sizes of \SI{4}{\nano\meter} x \SI{4}{\nano\meter} x \SI{0.8}{\nano\meter}, respectively. Fig.~\ref{fig:magsim} shows simulations for the bilayer flipping transition (Fig.~3 main text). We estimate the strain profile across the stack from the Raman measurements, and adjust the interlayer exchange coupling accordingly. We compare the behaviour in dependence of the applied field in the twisted and bare regions, both with and without exchange coupling across the twisted interface (for simplicity we take the pristine interlayer exchange coupling). This allows to assess whether the intraflake exchange interaction could be responsible for the observed low field forced-FM behaviour. The hysteresis loop is repeated ten times in order to generate some statistics. A few of these runs were taken out due to malfunctions in the simulation.

The onset of the flipping transition is indeed earlier than in the bare region in the simulations, but this behaviour does not appear to be linked to the inclusion of the exchange term. In fact, the hysteresis is narrower when the interflake exchange is added. These simulations indicate that interflake exchange plays a minor role in this case and structural defects likely dominate. 

\section{Additional data}

\subsection{Raman measurements}
\begin{figure}[htb]
    \centering
    \includegraphics[width=1\linewidth]{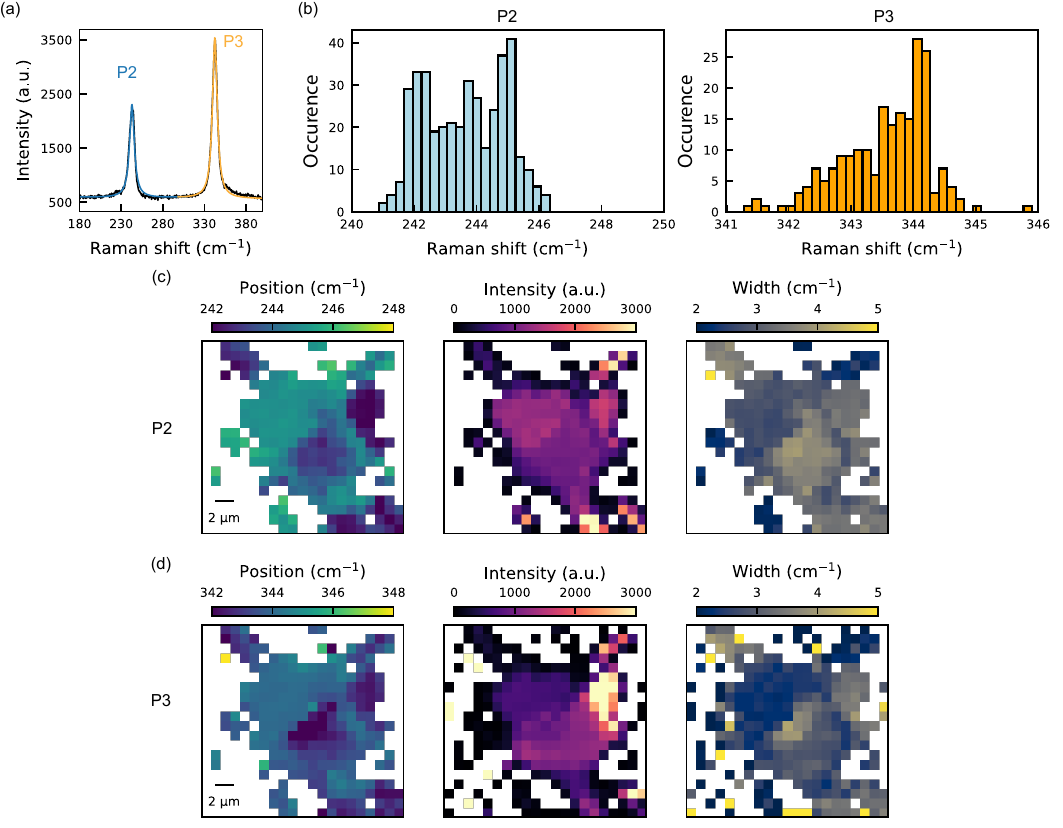}
    \caption{Spatial Raman measurements of the AFM-AFM stack. 
    \textbf{a} Exemplary Raman spectrum from a single pixel. 
    \textbf{b} Distribution of the shift of the P2 (left panel) and P3 (right panel) Raman lines for the AFM-AFM stack. 
    \textbf{c} Lorentzian fits results for the position, intensity, and width of the P2 Raman line on the AFM-AFM stack, where an intensity threshold was used to remove areas containing no CrSBr. 
    \textbf{d} Same as in \textbf{c} but for the P3 Raman line. 
    }
    \label{fig:raman}
\end{figure}

The Raman map shown in Fig.~3 was collected on a Horbia LabRAM Raman spectrometer using a 1800 lines/mm grating, a 532~nm Ventus laser and a 100$\times$ (NA=0.8) air objective resulting in a spot size of approximately 2~$\mu$m in diameter.  The laser power was kept to 2.8 mW (as measured before the objective) to avoid laser heating of damage of the sample. In Fig.~\ref{fig:raman} we present more detailed analysis of this data. A typical CrSBr Raman spectrum is shown in Fig.~\ref{fig:raman}(a), showing the P2 and P3 Raman lines. Both lines are modulated by strain and are observed to vary across the stack, as summarised by the histograms of fit peak locations Fig.~\ref{fig:raman}(b), where a Lorentzian lineshape was used for the fit. Maps of the peak locations, intensity, and width are shown in Fig~\ref{fig:raman}(c) for P2 and Fig.~\ref{fig:raman}(d) for P3. The features are broadly similar, however for the main text we focussed on the P3 line since it is less sensitive to the number of layers present in a stack.

The increased width in peaks fit within the top flake (the transition of which we analysed in Fig.~3 of the main text) highlights that there could be significant strain variation below the diffraction limit of the Raman measurement. 

\subsection{Other twisted monolayer transition}

\begin{figure}[htb]
    \centering
    \includegraphics[width=0.5\linewidth]{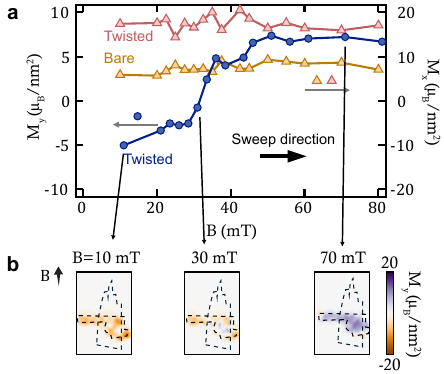}
    \caption{(a) Monolayer flipping data for the flake not shown in Fig. 1 of the main text. $M_y$ shows the transition and we also plot $M_x$ to show that it is unchanged both in the twisted region and in the bare (orthogonal) monlayer across this transition. (b) Images showing the switching transition, showing domain nucleation in the twisted region. }
    \label{fig:othermono}
\end{figure}

In Fig.~\ref{fig:othermono}(a) we show the switching transition of the monolayer not shown in the main text. The transition has step-like features but is generally smoother than that shown in the main text, occurs at lower fields ($<60$~mT), and over a wider span of field strengths (about 40~mT). 
The magnetisation reconstruction for this set of data is generally more prone to artefacts due to the aligned flake's small size compared to both the measurement spatial resolution and the thicker sections of the orthogonal flake. 
To circumvent the issue, for this set of recontructions we confine $M_y$ magnetisation to the target monolayer while allowing $M_x$ to be placed anywhere within the stack. We note that even with this procedure, the reconstructed magnetisation is lower that for a typical saturated monolayer, which is an artefact incurred by NV magnetometry when the magnetic domain size approaches the measurement spatial resolution. 
We can see in Fig~\ref{fig:othermono}(b), where select normalised magnetic images at extreme points in the transition are presented, that domains again appear to nucleate in (and are initially confined to) the twisted region.
However, due to the extremely small size of any ``bare" regions or regions that overlap with thicker twisted flakes, we do not include traces of their evolution in Fig.~\ref{fig:othermono}(a). 
Instead, we show $M_x$ (triangular points) averaged over the same regions as in the main text, showing that the magnetisation in the other flake appears unchanged across this transition. 

The broad and reproducible switching transition observed in this case allows us to consider a final experiment, wherein we initialise one flake (``primed") into a multi-domain state using a field of $\approx 35$~mT  and sweep across the switching transition of the orthogonal flake (``unprimed"). This experiment allows us to further test the hypothesis that domain walls in one flake can influence (or be influenced by) magnetisation in the other due to the local spin (anti)alignment. 

\begin{figure}[htb]
    \centering
    \includegraphics[width=0.5\linewidth]{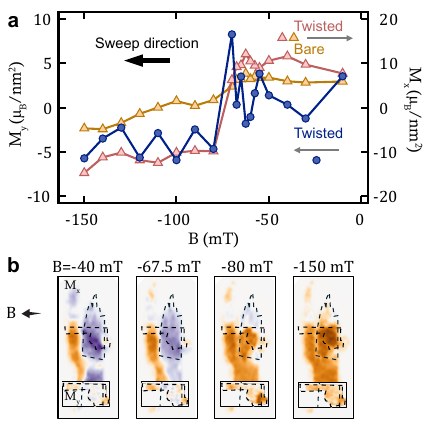}
    \caption{(a) Summary of the FM switching transition sweeping along the $y$ direction with the orthogonal (primed) flake initialised in a partially-switched state. (b) Select images from the series in (a), showing $M_x$ with $M_y$ inset in all cases. }
    \label{fig:perpinit}
\end{figure}

In this case (data summarised in Fig.~\ref{fig:perpinit}(a)) we see a sharp transition in the twisted region of the unprimed flake (given by the $M_x$ magnetisation), which is accompanied by the $M_y$ magnetisation from the primed flake saturating as we can see in the images Fig~\ref{fig:perpinit}(b).  Compared to the data shown in Fig.~1 of the main text, the FM switching transition onset in this case is later (80~mT versus 67.5~mT). While the difference may simply be due to the stochastic nature of domain nucleation, it may also hint at a relative increase in energy cost for domain nucleation in the unprimed flake due to the scrambled state of the primed flake. We also do not observe the domain border alignment with the $a$ ($b'$) axis within the twisted region that we saw in Fig.~1. The small flake sizes relative to our measurement spatial resolution and the potential impact of conflating variables such as stacking-induced strain do not allow us to make definitive conclusions in this case, however these results, combined with those shown in Fig.~4, are suggestive of a non-trivial magnetic interaction across the twisted interface. 
\subsection{Other twisted bilayer transition}

\begin{figure}[htb]
    \centering
    \includegraphics[width=\textwidth]{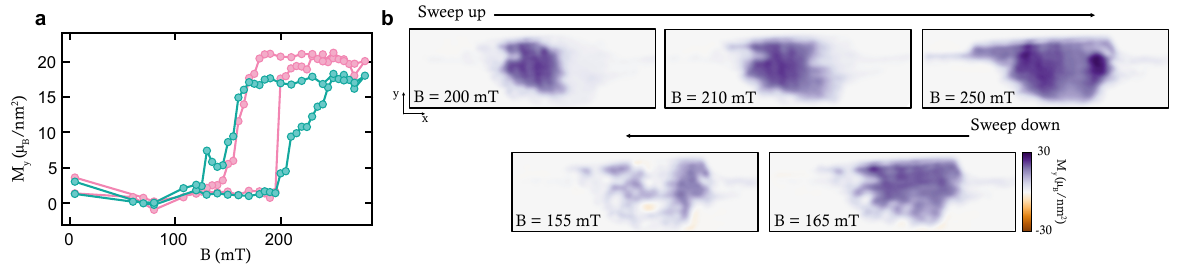}
    \caption{(a) Hysteresis loop for the bilayer transition observed when sweeping fields along the $y$ direction. Twisted region shown in pink, bare region in cyan. (b) Exemplary images from the measurement series in (a). }
    \label{fig:bibi}
\end{figure}
In Fig.~\ref{fig:bibi}(a) we show the hysteresis observed when sweeping the applied magnetic field along the $y$ axis. In this case, we do not observe strong magnetic signals at low field or an offset hysteresis as for the other bilayer phase transition. Instead, a narrower hysteresis loop is observed in the twisted region, similar to the results in Fig.~2 of the main text. The difference between the two behaviours within the same twisted stack hints at factors other than exchange interactions across the twisted interface being important in determining the magnetic properties. Namely, as the top flake in this stack was observed to be more highly strained (c.f. Fig.~3(b)), it is likely that the extra level of strain plays a role in enabling some of these behaviours.  

Along with the narrower hysteresis, we again observe that the strength of the magnetisation in the twisted region is stronger after the phase flip than in the bare region. The increase in magnetisation is sudden and only present during the forced FM state, meaning it is not easily explained by a gradual canting of the misaligned flake. 

\subsection{AFM switching transition in AFM-FM stack}
\begin{figure}
    \centering
    \includegraphics[scale=1]{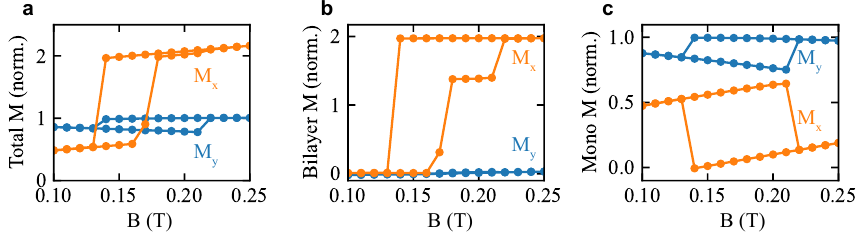}
    \caption{(a) Results of a micromagnetic simulation of the experiment from Fig.~2 of the main text. The net magnetisation in the twisted region (modelled as a 2L-1L system for computational simplicity) is calculated and decomposed into $M_x$ (orange) and $M_y$ (blue) components. The magnitude is normalised against the magnetisation corresponding to a saturated CrSBr monolayer. (b) As (a) but only for the bilayer. (c) as (a) but only for the FM layer.}
    \label{fig:AFMFM_sims}
\end{figure}

In Fig.~2 of the main text we saw that the AFM switching transition in an AFM-FM stack onset earlier in the twisted region, and exhibit a narrower hysteresis loop. In Fig.~\ref{fig:AFMFM_sims} we present results from micromagnetic simulations of this stack, modelled as a bilayer-monolayer system for computational simplicity. Figure~\ref{fig:AFMFM_sims}(a) shows the total simulated magnetisation in the twisted region, with the projection $M_x$ seen to closely resemble the data from Fig.~2(d). Notably, the net magnetisation is nonzero before the bilayer flipping transition and is above the value for a saturated bilayer afterwards. Figure~\ref{fig:AFMFM_sims}(b) and (c) show the contributions just from the bilayer and monolayer, respectively. Here we can see that the FM layer cants steadily towards the $x$ axis with increasing field and, if nonzero exchange is assumed, acts to inhibit the full flipping of the bilayer. When the bilayer flips fully, the corresponding increase in $M_x$ is almost totally offset [c.f. Fig.~\ref{fig:AFMFM_sims}(a)] by a reduction in monolayer $M_x$. Upon decreasing the field after saturation, we see that the monolayer hysteresis matches the width of that of the bilayer transition. 

\begin{figure}[htb]
    \centering
    \includegraphics[scale=1]{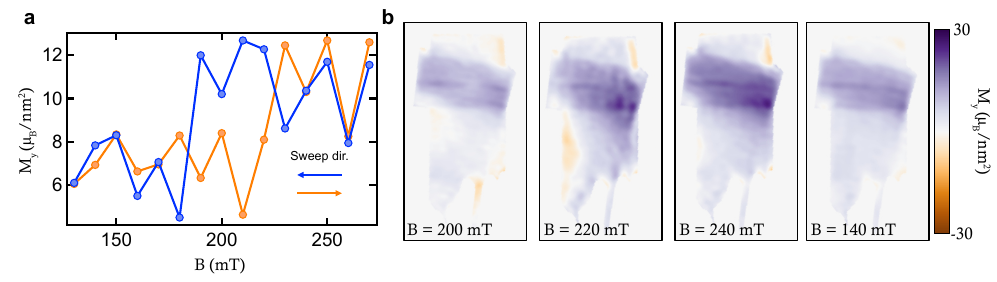}
    \caption{(a) $M_y$ hysteresis loop observed for the data from the sweep shown in Fig.~2 of the main text. The values plotted are averages from within the twisted region, with the average from the bare region subtracted. (b) Selected $M_y$ images from this series.}
    \label{fig:My_AFMFM}
\end{figure}

Since the contributions from individual layers cannot be measured in our experiment, we are unable to directly confirm the simulated picture. However, we can turn to the reconstructed $M_y$ component from the experimental dataset as we can see from Fig.~\ref{fig:AFMFM_sims}(c) that any hysteresis in monolayer spin reorientation is evident in both magnetisation projections. In Fig.~\ref{fig:My_AFMFM}(a) we plot the average reconstructed $M_y$ magnetisation in the twisted region, normalised against the bare bilayer. A positive magnetisation from the FM layer is always observed, however an increase is observed across the AFM switching transition (example images shown in Fig.~\ref{fig:My_AFMFM}(b)). Unlike in the simulations, the observed $M_y$ hysteresis is narrower than (and slightly offset from) the bilayer transition. Nevertheless, these results do appear to suggest that the FM layer reorients across the bilayer flipping transition. As in the simulations, a nonzero exchange interaction across the twisted interface is a candidate cause, although it is also difficult to rule out confounding factors such as strain.



\bibliographystyle{naturemag}
\bibliography{bib.bib}